# A Secure Cyclic Steganographic Technique for Color Images using Randomization

K. Muhammad[1], J. Ahmad[2], N. U. Rehman[3], Z. Jan[4], R. J. Qureshi[5]

[1,2,3,4,5] Department of Computer Science, Islamia College Peshawar, K.P.K, Pakistan
Khan.muhammad.icp@gmail.com

***Abstract*-**Information Security is a major concern in today's modern era. Almost all the communicating bodies want the security, confidentiality and integrity of their personal data. But this security goal cannot be achieved easily when we are using an open network like Internet. Steganography provides one of the best solutions to this problem. This paper represents a new Cyclic Steganographic Technique (CST) based on Least Significant Bit (LSB) for true color (RGB) images. The proposed method hides the secret data in the LSBs of cover image pixels in a randomized cyclic manner. The proposed technique is evaluated using both subjective and objective analysis using histograms changeability, Peak Signal-to-Noise Ratio (PSNR) and Mean Square Error (MSE). Experimentally it is found that the proposed method gives promising results in terms of security, imperceptibility and robustness as compared to some existent methods and vindicates this new algorithm.

***Keywords*-**Steganography, Least Significant Bit, Cyclic Steganographic Technique

## I. INTRODUCTION

Steganography is a Greek origin word meaning "Concealed Writing". It can be considered as a way to hide secret information in cover image pixels such that it cannot be detected by Human Visual System (HVS) and nobody know about its existence without the intended sender and receiver. Steganography requires three main components named as carrier object, secret data and steganographic algorithm. Sometimes a secret key and cryptographic algorithm is also required in order to increase the security levels and introduce multiple barriers in the way of an attacker. Steganography can be used for many useful applications like online voting security, secure transmission of top-secret data between national and international governments, online banking security, military and intelligent agencies security and safe circulation of secret documents among defense organizations. On the other hand, Steganography is also very nefarious; it is used by terrorists and criminals for their secure communication and sending viruses and Trojan horses to compromise machines. [1-4].

*1.1 Types of Steganography w.r.t Carrier Object*

There are five different types of steganography based on the carrier object that is used for embedding the secret information. The carrier object may be images, text, videos, audios or network protocol packets. If the image is used as a carrier, it is called image steganography. Similarly if video is used for hiding secret messages, we call it video steganography and so on[1, 5]. The diagrammatic representation of different types of steganography is shown in Fig. 1. The types of steganography are:

a. Audio Steganography
b. Image Steganography
c. Video Steganography
d. Text Steganography
e. Network Steganography

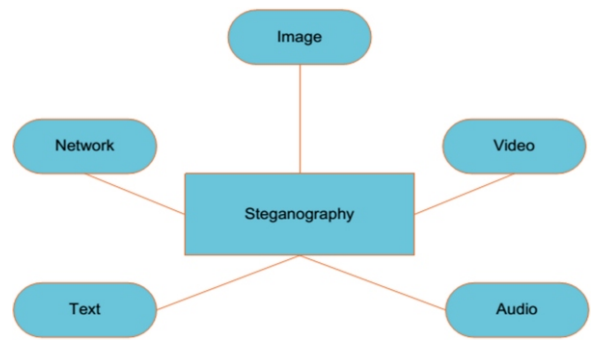

Fig. 1. Types of Steganography w.r.t carrier object

*1.2 Types of Steganography w.r.t Key Exchange*

To make the steganographic algorithm more robust against different fraudulent behaviors, the idea of using secret key and public key was introduced. When secret key and public keys are used in steganography, a mechanism should be there via which these keys can be securely exchanged. Usually one or two keys are used in the process of steganography; public key that is used to embed secret information into the carrier object and secret or private key that has a mathematical relationship with public key and is used for extraction of secret data from stego object [6]. On the basis of exchange of keys, there are three types of steganography as shown in Fig. 2.

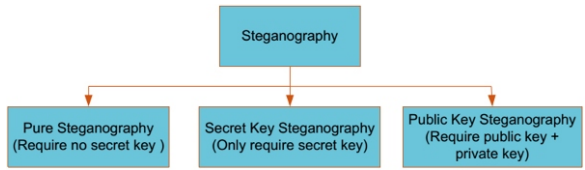

Fig. 2. Types of steganography w.r.t Key Exchange

*1.3 Classifications of Steganographic Techniques*

The classification of steganographic techniques can be performed using different approaches. One approach is to classify them on the basis of carrier object that is used at the time of embedding the secret data into the carrier object. Another approach is to categorize them on the basis of cover modification in the process of hiding secret information. We use the second approach and classify steganographic techniques into two broad domains.

*1.3.1 Spatial Domain Techniques*

In spatial domain techniques, the carrier object (image, video etc) pixels are directly changed in order to hide secret data inside it. These techniques have high payload and bring minor changes in the carrier object but are vulnerable to even simple statistical attacks like cropping, scaling, rotating, compression etc. Some of the techniques that belong to spatial domain are:
a.  Least Significant Bit (LSB)
b.  Gray-Level Modification (GLM)
c.  Pixel Value Differencing (PVD)
d.  Edges based Embedding (EBE)

*1.3.2 Transform Domain Techniques*

In transform domain techniques, the carrier object (image, video etc) is first transformed from spatial domain to transform domain and then its frequencies are used to hide the secret data. After embedding the secret data, the object is again transformed into spatial domain. These techniques have lower payload but are robust against statistical attacks. Some techniques of transform domain are[7]:
i.   Discrete Wavelet Transform Technique (DWTT)
ii.  Discrete Fourier Transform Technique (DFTT)
Iii. Discrete Cosine Transform Technique (DCTT)

In this paper, color image has been selected as a carrier object because it contains more redundant bits and color image provide more pixels to increase the payload capacity. The remaining of the paper is organized as fellows. Section 2 describes some existence related approaches whose limitations led us towards current proposed work. Section 3 briefly discusses the proposed technique. Section 4 discusses experimental results and discussion and section 5 concludes the paper.

## II. LITERATURE REVIEW

Steganography was first started by ancient Greeks in (484-425 BC). Histaeus, the ruler of Miletus shaved the head of his honest slave and then tattooed a secret text or a symbol on his head. When the slave's hair grew back, the slave was sent to Aristagorus in order to provide him the hidden text. When the slave arrived at his destination, his head was shaved again in order to read the secret message. From that time to till now many technical, linguistics and modern steganographic techniques have been developed and used for steganography. All the techniques have their corresponding pros and cons. Some techniques have high payload capacity and good imperceptibility depending upon the selected cover for secret data hiding (Spatial domain techniques) but more vulnerable to attacks (Noise throwing, cropping, rotation, resizing etc) while others techniques are more robust against statistical attacks but they have lower payload capacity. This means that there is always a tradeoff between the three factors (Payload, Imperceptibility and Robustness). As an example a few techniques of steganography are critically discussed below [8-10].

The most basic technique used for steganography is LSB technique in which the least significant bit of carrier image pixels are replaced with the bits of secret message. To clarify the concept of LSB method, consider the pixels below and hide a secret character (S.C) "B" inside it.

| Decimal | Binary | S.C | Binary |
|---------|----------|-----|----------|
| 143 | 10001111 | B | 01000010 |
| 134 | 10000110 | | |
| 126 | 01111110 | | |
| 99  | 01100011 | | |
| 44  | 00101100 | | |
| 134 | 10000110 | | |
| 79  | 01001111 | | |
| 127 | 01111111 | | |

Now replace the LSBs of the given pixels with the secret message bits as shown.

| Decimal | Binary |
|---------|----------|
| 142 | 1000111**0** |
| 135 | 1000011**1** |
| 126 | 01111110 |
| 98  | 0110001**0** |
| 44  | 00101100 |
| 134 | 10000110 |
| 79  | 01001011 |
| 126 | 0111011**0** |

The bold face LSBs are changed which shows that approximately half of the pixels changes so the resultant stego image slightly changes from the original cover image.

In [11], the authors propose a robust method which embeds variable bits in image pixels depending on the pixel value and value of mean and standard deviation (SD). Two bits are inserted in the image pixel if (meanSD/2) is greater than pixel value; 3 bits are stored if (mean +SD/2) is greater than pixel value otherwise 4 bits are stored in each pixel value. Chaotic effect is also obtained in the proposed method by using random traversing path which make the attack loathsome but nothing is given about generating the random traversing path which is its major weak point.

Reference [12] presents pixel indicator technique (PIT) in which one channel is used for indication while other two channels are used for embedding secret data in a predefined cycle manner which enhances the robustness of proposed method. The experimental results demonstrate the larger payload and enhanced imperceptibility of the proposed method. This method also eliminates the stego key exchange overhead. The major limitation of this technique is the fixed number of bits embedded in each gray level of the host image which may cause noticeable distortion if we increase the number of embedded bits. Furthermore, the payload of this method is absolutely dependent on the carrier image and indicator bits which may be reduced.

In [13], the authors proposed a new method to embed secret data in the GREEN or BLUE channel of carrier image on the basis of secret key bits and RED channel LSB. This method adds one more level security to the existing LSB method by utilization of secret key. The RED channel LSB and secret key bit is xored and then a decision is taken on the basis of its result to replace the LSB of GREEN or BLUE channel. The proposed method has the same payload, more robustness and better security as compared to simple LSB method. However the secure key exchange of secret key is an open challenge and is an extra overhead of proposed method.

## III. PROPOSED METHOD

In this paper, a more secure steganographic technique is presented which hides secret data in the LSBs of cover image pixels in a randomized cyclic manner. The order in which secret bits are embedded in cover image pixels' planes is RED, GREEN, BLUE, RED, GREEN, and BLUE and so on. This randomized and cyclic approach increases the robustness of the proposed algorithm and randomly disperses the secret data inside the cover image pixels. Due to this reason it is difficult for a malicious user to extract the original secret data from the stego image. The embedding and extraction algorithm for the proposed method are given below.

*3.1 Embedding Algorithm*
*Input:* Color Image and secret data
*Output:* Stego Image
Step 1: Take the cover color image and secret data.
Step 2: Separate the RED, GREEN and BLUE planes from the cover image.
Step 3: Convert secret data into 1-D array of bits.
Step 4: Set channelFlag = 1 initially (channelFlag determines the channel for embedding).
Step 5: If channelFlag = 1
Replace the LSB of RED channel with secret bit Else if channelFlag = 2
Replace the LSB of GREEN channel with secret bit Else if channelFlag = 3
Replace the LSB of BLUE channel with secret bit End
Step 6: Increment channelFlag by 1.
Step 7: If channelFlag = 3
Set channelFlag = 1;
End
Step 8: Repeat Step 5 to Step 7 until all secret data bits are embedded.
Step 9: Combine all three planes to form the resultant stego image.

<Fig. 3: Embedding Algorithm Flowchart >

*3.2 Extraction Algorithm*
*Input:* Stego Image
*Output:* Secret data
Step 1: Take the stego image and separate the RED, GREEN and BLUE planes from it.
Step 2: Set channelFlag = 1 initially.
Step 3: If channelFlag = 1
Extract the LSB of RED channel.
Else if channelFlag = 2
Extract the LSB of GREEN channel.
Else if channelFlag = 3
Extract the LSB of BLUE channel.
End
Step 4: Increment the channelFlag by 1.
Step 5: If channelFlag = 3
Set channelFlag = 1;
End
Step 6: Repeat Step 3 to Step 5 until all secret data bits are extracted.
Step 7: Convert the extracted secret bits into its original secret data format.

<Fig. 4: Extraction Algorithm Flowchart >

## IV. EXPERIMENTAL RESULTS AND DISCUSSION

The proposed algorithm, LSB algorithm and Karim's algorithm[13] are simulated using MATLAB R2013a. The standard color images used for experiments of implemented three techniques are Lena, baboon, trees etc. Experimentally, these three algorithms are evaluated by three different perspectives:
a. Embedding same amount of cipher in different standard color images of same dimensions.
b. Hiding same amount of secret data in the same

Image of different dimensions.
c. Embedding variable amount of cipher in the same image of same dimensions.

*4.1 Comparison of proposed method and existing methods*

The comparison of proposed technique with existing techniques is based on two types of analysis known as subjective and objective analysis. Subjective analysis is done using Human Visual System (HVS) to notice the changes between the cover and stego images and their corresponding histograms. A few samples of standard color cover and stego images and their histograms for the proposed method are shown below in Fig. 5-7. From figures it is observed that there is no noticeable change in the cover and stego images and their histograms which shows the effectiveness of the proposed method.

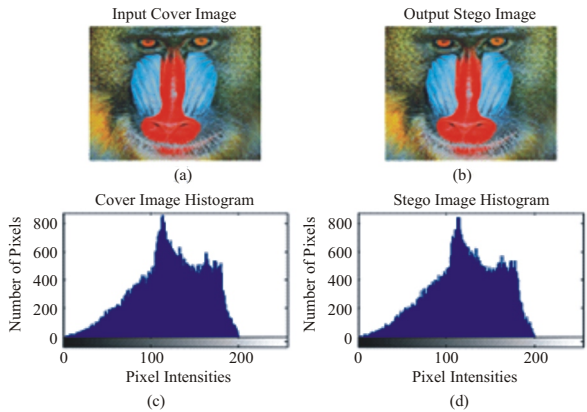

Fig. 5. baboon cover and stego image and their histograms

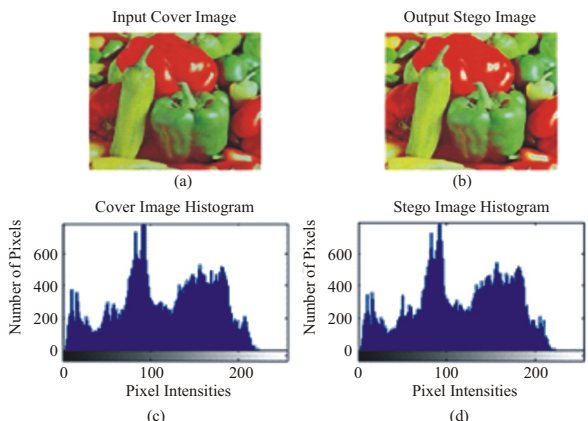

Fig. 6. Peppers cover and stego image and their histograms

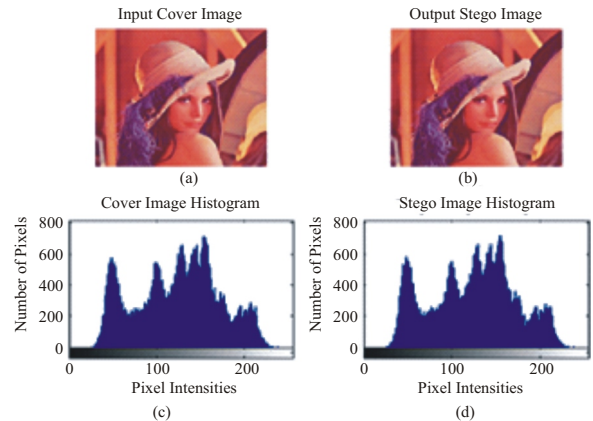

Fig. 7. Lena cover and stego image and their histograms

Objective analysis is the second mathematical standard for measuring the distortion that occurs in the cover image after embedding secret data. Objective analysis is performed on the proposed method using Peak Signal-to-Noise Ratio (PSNR) and Mean Square Error (MSE). PSNR and MSE are calculated using equations (1) and (2).

$$PSNR = 10 log_{10}\left(\frac{C_{max}^2}{MSE}\right) \quad (1)$$

$$MSE = \frac{1}{MN}\sum_{x=1}^{M}\sum_{y=1}^{N}(S_{xy} - C_{xy}) \quad (2)$$

Here M and N are image dimensions, x and y are loop variables, S is stego image, C is cover image and $C_{max}$ is the maximum pixel intensity among both images. The experimental results of the proposed method, LSB and method in [13] are shown in Table I, Table II and Table III respectively.

TABLE I
COMPARISON BASED ON PSNR

| Image Name | LSB Method | Karim's Method [13] | Proposed Method |
|---|---|---|---|
| | PSNR (dB) | PSNR (dB) | PSNR (dB) |
| baboon.png | 61.8784 | 48.558 | 52.5573 |
| lena.png | 42.6331 | 42.6204 | 71.8865 |
| building.png | 51.4677 | 47.0305 | 63.4042 |
| parrot.png | 49.708 | 49.8421 | 49.8417 |

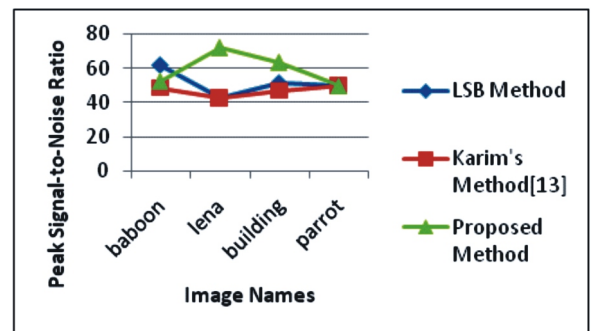

Fig. 8. Comparison based on PSNR with different images of dimension 256×256

TABLE II
PSNR BASED COMPARISON OF EXISTING AND PROPOSED METHOD

| Image Dimensions | LSB Method PSNR(dB) | Karim's Method [13] PSNR(dB) | Proposed Method PSNR(dB) |
|---|---|---|---|
| 128×128 | 70.3187 | 65.5328 | 65.5474 |
| 256×256 | 61.8784 | 50.8811 | 52.5573 |
| 512×512 | 52.4555 | 37.2456 | 48.8123 |
| 1024×1024 | 59.204 | 41.9577 | 49.7897 |

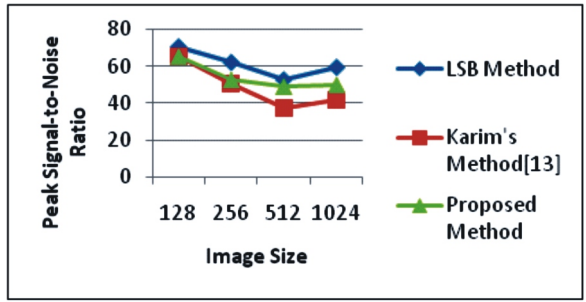

Fig. 9. PSNR based comparison with same size cipher & different image dimensions

TABLE III
COMPARISON BASED ON PSNR WITH VARIABLE AMOUNT OF CIPHER EMBEDDED

| Image Name | Cipher size in (KBs) | LSB Method | Karim's Method [13] | Proposed Method |
|---|---|---|---|---|
| | | PSNR(dB) | | |
| baboon with dimension 256×256 | 2 | 63.3775 | 52.0373 | 52.0668 |
| | 4 | 61.8442 | 51.6345 | 51.6853 |
| | 6 | 60.4909 | 51.1776 | 51.2539 |
| | 8 | 59.7481 | 50.8811 | 51.0035 |

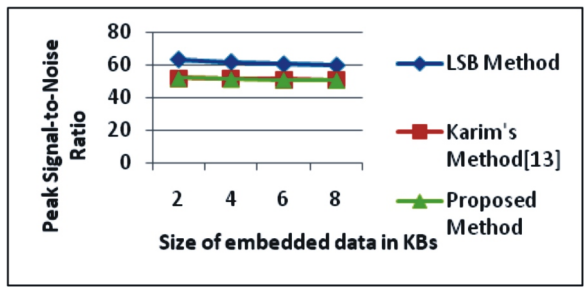

Fig. 10. Comparison based on PSNR with same image dimension and variable amount of cipher

*4.2 Performance Analysis of the proposed method*

The performance of a steganographic technique is measured using three metrics; payload, robustness and imperceptibility. An algorithm is considered to be best if it has high imperceptibility, more robustness and larger payload. The payload of all three mentioned approaches is same i.e. 1bpp (bits per pixel). The proposed scheme is better than classical LSB scheme in terms of robustness because the classical LSB method directly hides data in only blue channel which can be easily detected while the proposed method scatters it in all three channels and hence increases the robustness. In contrast to Karim's method, the proposed method gives better imperceptibility as indicated by larger PSNR values in Table 1-3.

*4.3 Advantages and limitations of the proposed method*

The proposed method embeds the secret data inside the gray levels of the host image in a randomized cyclic manner which increases its robustness and makes the extraction difficult. Furthermore, the proposed scheme gives better imperceptibility as compared to existing methods which can be also confirmed by larger values of PSNR. The major weakness of the proposed method is its vulnerability to different image processing and statistical attacks such as image cropping, scaling and noise attacks. Since spatial domain is used for the proposed scheme, therefore the embedded data will be lost like other schemes if the stego image is compressed, rotated or attacked with different types of noises (pepper noise, salt and pepper, speckle noise etc).

## V. CONCLUSIONS

In this paper, we proposed a more secure cyclic steganographic algorithm for RGB images using the concept of randomization with enough robustness, imperceptibility and security. An average PSNR above 50dB is achieved by the proposed algorithm which describes its superioty as compared to some existing steganographic algorithms. The proposed method hides the same amount of secret data as LSB and Karim's method but scatters it inside the whole image pixels to increase the robustness. The utilization of randomization and cyclicness makes the extraction of original secret information from the stego image more difficult for a malicious user. Hence the proposed technique gives promising results in terms of robustness, imperceptibility and security.

## VI. ACKNOWLEDGMENT

The authors wish to thank all the contributors for their critical and technical review of the proposed work and their valuable support and guidance. Special thanks to Mr. Zahid Khan for his valuable help and support during this research work

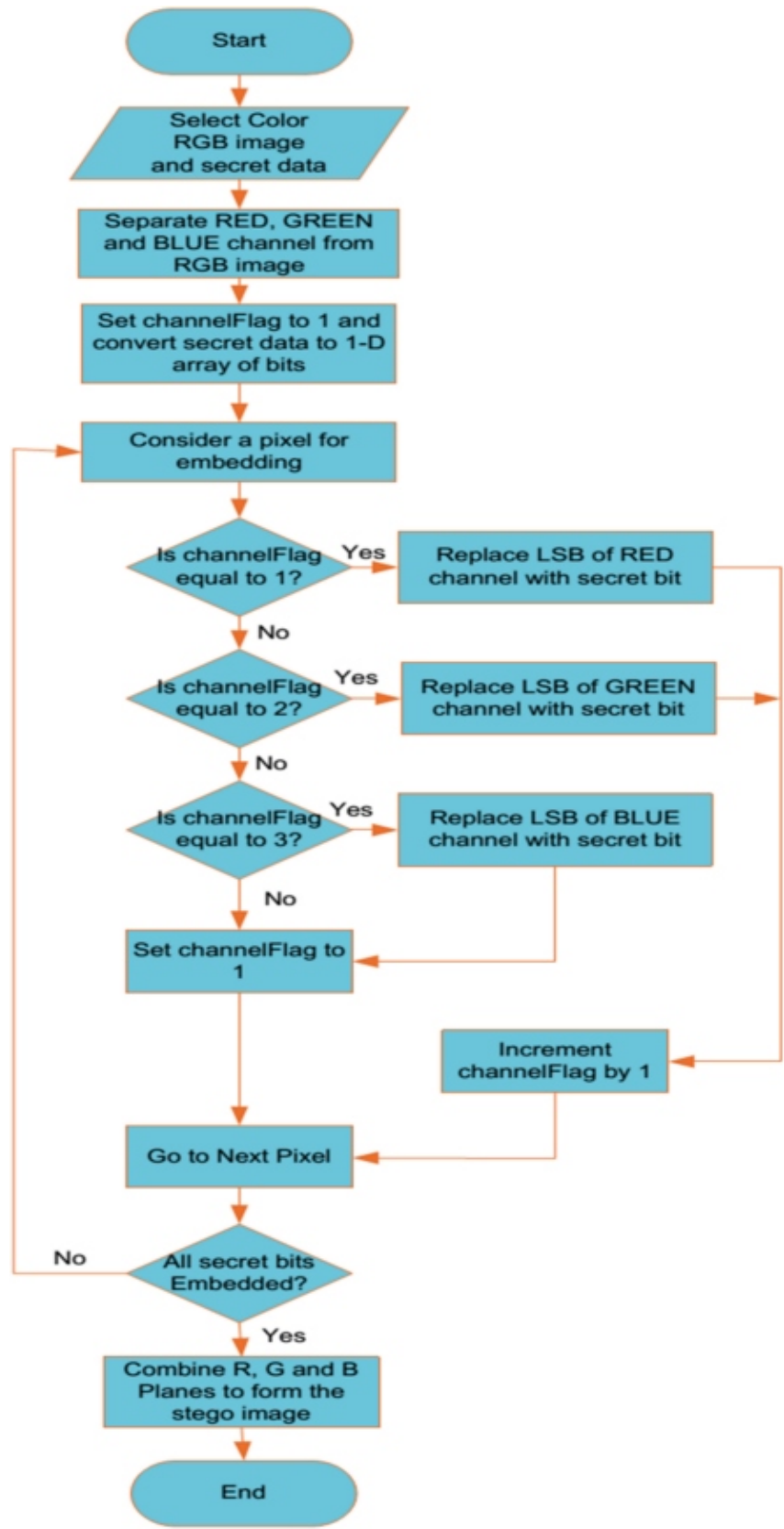

Fig. 3. Embedding Algorithm Flowchart

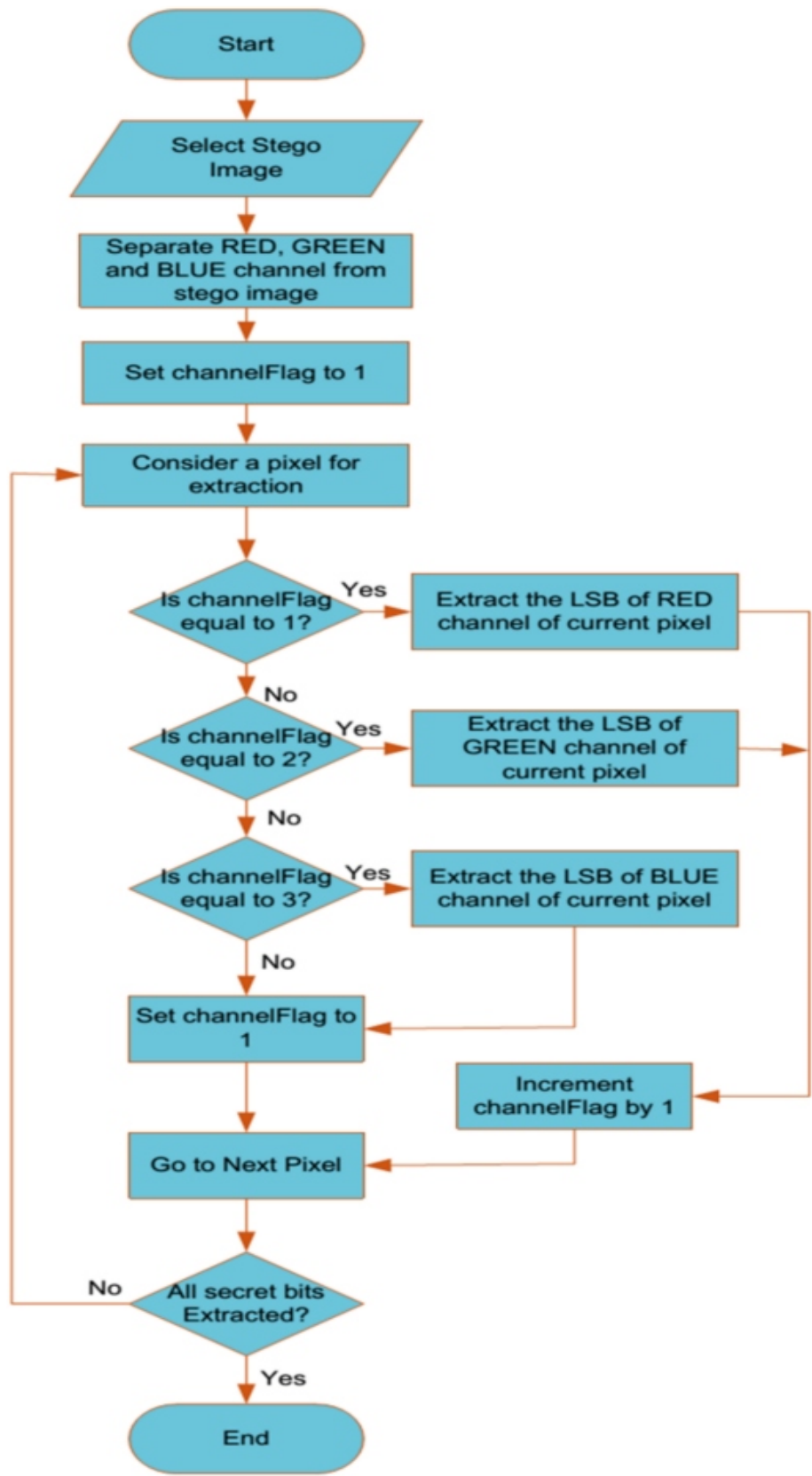

Fig. 4. Extraction Algorithm Flowchart